Comment: LaTeX file
-------------------------------------------

\def\truelbl#1\mbox{\rlap{\hbox to \marginparsep{\hfil}\fbox{(\small #1)}}}

\def\mn#1{\marginpar[{#1}]{#1}}

\def\mf#1{\protect\mn{\footnotesize#1}}
\def\msz#1{\protect\mn{\scriptsize#1}}

\def\mn#1{}
\def\mf#1{}
\def\msz#1{}

\def\meq#1{\mf{(#1)}}
\def\meq#1{}

\newcommand{\sch}{Schr\"odinger}

 \newcommand{\bpartial}{{{\partial}_{\sbf x}}}

 \newcommand{\sbf}[1]{\mbox{\scriptsize$\bf{#1}$}}

\newfont{\twelvembsy}{cmbsy10 scaled1200}
\newfont{\eightmbsy}{cmbsy10 scaled800}
\newfont{\sixmbsy}{cmbsy10 scaled600}

\newfont{\twlmib}{cmmib10 scaled1200}
\newfont{\egtmib}{cmmib10 scaled800}
\newfont{\sixmib}{cmmib10 scaled600}

\documentstyle[12pt]{article}

\title{\vspace{-2.5cm}
 Quantum Equivalence Principle \\ for
 Path Integrals \\ in Spaces with Curvature and Torsion%
\thanks{Work supported in part by Deutsche Forschungsgemeinschaft
under grant no. Kl.~256.
}
 }

\author{\vspace{1cm}H. Kleinert
        \\[-0.5cm]
        {\normalsize  Institut f\"{u}r Theorie der Elementarteilchen} \\[-5pt]
        {\normalsize  Freie Universit\"{a}t Berlin} \\[-5pt]
        {\normalsize  Arnimallee 14 \ \ \ \ \ D - 1 Berlin 33}
       }
{{\date{
}}}

\begin{document}

\maketitle
\vspace{-.5cm}
\begin{abstract}
We formulate a new quantum equivalence principle by which
a path integral for a particle in a general metric-affine space is
obtained from that in a flat space by a non-holonomic
coordinate transformation.
The new path integral is free of the ambiguities
of earlier proposals and the ensuing Schr\"odinger equation
does not contain the often-found but
physically false terms proportional to the scalar curvature.
There is no more quantum ordering problem.
For a particle on
the surface of a sphere in $D$ dimensions, the new path integral
gives the correct energy $\propto \hat L^2$ where $\hat L$ are the
generators of the rotation group in ${\bf x}$-space. For the transformation of
the
Coulomb path integral
to a harmonic oscillator, which passes at
an intermediate stage a space with torsion,
the new path integral
renders the correct energy spectrum with no unwanted time-slicing corrections.
\end{abstract}

\newpage
%
%
\noindent
1)~In Schr\"odinger quantum mechanics, the dynamical properties
 of a point particle
are governed by the
differential equation $\hat H\psi ({\bf x},t)=i\hbar \partial _t\psi ({\bf
x},t)$
where $\psi ({\bf x},t)$ is a square-integrable probability amplitude.
If the particle moves in flat space parametrized by
cartesian coordinates, the Hamilton operator $\hat H$ is
found from the {\em correspondence
principle\/}. It prescribes that $\hat H$
be obtained from the classical Hamiltonian $H({\bf p},{\bf x})$ by replacing
the
canonical coordinates $x^i$ and momenta $p_i$
by the \sch{} operators
$\hat x^i=x^i$, $\hat p_i=-i\partial /\partial x^i$.
If the
particle moves on the surface of a sphere in $ D$ dimensions
or if the system is
a spinning top, things are not
as simple. Then
canonical quantization rules are
of no help due to ordering ambiguities
arising in the attempt to do the above operator
replacement
in the kinetic part of the Hamiltonian, which now has the general form
$$
 H=\frac{1}{2M} p_\mu g^{\mu \nu }(q) p_\nu $$
with no principle instructing us how to order the momentum operators $\hat
p_\mu =-i\partial /\partial q^\mu $
with
with respect to the position variables $\hat q^\mu $.

For a particle on the surface of a sphere the
most obvious way out of this dilemma is to make use of the
symmetry of the system, express the classical
Hamiltonian as a square of the classical angular momentum ${\bf L}$,
$$H=\frac{1}{2Mr^2}{\bf L}^2$$
($r=$ radius of the sphere), and replace ${\bf L}$ by the
differential operators  $\hat {\bf L}$ which generate rotations
in the space of
square integrable functions on the surface.
The resulting Hamilton operator turns out to coincide with
\begin{equation} \label{}
\hat H =-\frac{1}{2Mr^2} \Delta
\label{so}\end{equation}
where $\Delta $ is the
Laplace-Beltrami operator
\begin{equation} \label{1.labe}
\Delta=\frac{1}{\sqrt{ g}}\partial _\mu g^\mu {}^\nu \sqrt{ g}\partial _\nu .
\end{equation}
This agrees with what we would find from the flat-space Hamiltonian operator
\begin{equation} \label{}
\hat H=-\frac{1}{2M}\bpartial ^2
\label{}\end{equation}
by subjecting it to a local {\em non-holonomic\/}
transformation $dx^i\rightarrow e^i_\mu (q)dq^\mu $ from flat cartesian
coordinates to
curved space.
This is
a nontrivial observation since initially one is allowed to transform
the flat-space Hamilton operator
only to new coordinates, which is always done
holonomically.

If one wants to find the correct
rules for path integration in curved spaces
one better comes out in agreement with such
group quantization rules. This has recently been achieved \cite{curtor,pi}.

In spaces with torsion
there is no physical system which could directly
be used for an experimental confirmation of the result.
Fortunately, however, there is the path integral of the hydrogen atom
\cite{dk}.
It can be defined and
solved only by a transformation to new coordinates in which
it becomes harmonic.
This transformation happens to be non-holonomic
and leads, at an intermediate stage, through a space with
torsion.  Also here it turns out that the correct
Hamiltonian operator at the intermediate stage
is simply
the non-holonomically transformed flat-space operator.

The path integral of the Coulomb system
can therefore be used as a testing ground for the
correct treatment of
torsion.

The recently proposed rule
\cite{curtor,pi}
for writing down a path integral in spaces with curvature and torsion
treats the above systems correctly.
It can therefore be considered as a
reliable {\em quantum equivalence principle\/} at the level
of path integrals telling us
how to generalize the Feynman path integral formula in
cartesian coordinates to non-euclidean spaces. Hopefully, it will
eventually lead to
the correct measure for the functional integral
of quantum gravity.
Earlier proposals for the path integral given by DeWitt \cite{dw}
and others \cite{ch,gav}
produced various additional constants
to the Schr\"odinger operator (\ref{so}) which were proportional
to the Riemannian scalar curvature $ \bar R$. For
a sphere of radius $ r$
this is $ \bar R=(D-1)(D-2)/r^2$ and for the top $3/2I$, where
$ I$ is the moment of inertia \{for the asymmetric top with three moments of
inertia it is
$[{(I_1+I_2 +I_3 )^2-2(I_1^2+I_2^2 +I_3^2)}]/
       {2I_1 I_2 I_3 }   $\}.
Apart from contradicting the above natural quantization via
the rotational Lie algebra such constants, if really present,
would change
the gravitational properties of
interstellar gases of rotating molecules and will be rejected.

The first important success of the new
quantum equivalence principle
was the closing of an outstanding gap
in the solution of the path
integral of the $D=3$ Coulomb system \cite{curtor}.
The previous solution
had been
only structural in character \cite{dk} and the proper treatment of
the time sliced expression had been limited only to the
unphysical
case of $ D=2$ dimensions \cite{howtodo},
the reason being that the combined coordinate and time
transformations which make the system harmonic and integrable
are holonomic in $D=2$ and do not
produce curvature nor torsion. In $D=3$ where this happens
it was the new quantum equivalence principle which finally led to
the solution \cite{hi}.

The purpose of this lecture is to review the
essence of this new
path integral approach.\\ ~ \\
2)~Our starting point is the certainly valid path integral in a
flat space parametrized with
euclidean coordinates ${\bf x}$. It has the time-sliced
form:
\begin{equation} \label{21}
({\bf x},t \vert {\bf x}',t') =
\frac{1}{\sqrt{2\pi \epsilon\hbar/M}^D}\prod_{n=2}^{N+1}
\left[
\int_{-\infty}^{\infty} d\Delta x_n \right] \prod_{n=1}^{N+1} K_0^\epsilon
(\Delta
{\bf x}_n),
\end{equation}
with the short-time amplitudes
\begin{equation} \label{22}
\!K_0^\epsilon (\Delta {\bf x}_n) =
\langle {\bf x}_n \vert\exp
\left\{ -\frac{1}{\hbar} \epsilon \hat{H}\right\}
\vert {\bf x}_{n-1}\rangle
\equiv \frac{1}{\sqrt{2\pi \epsilon \hbar/M}^D}
\exp\left\{{-\frac{1}{\hbar}\frac{M}{2}\frac{(\Delta{\bf
x}_n)^2}{\epsilon}}\right\}
\end{equation}
where $\Delta {\bf x}_n \equiv {\bf x}_n -{\bf x}_{n-1}, {\bf x}
\equiv {\bf x}_{N+1},
{\bf x}' \equiv {\bf x}_0$
(we may omit a possible extra
potential which would enter trivially).
We now transform $(\Delta {\bf x}_n)^2$
to a space with curvature and torsion
by a {\em non-holonomic\/} mapping \cite{gfcm}, parametrized with coordinates
$q^\mu$.
For infinitesimal $\Delta {\bf x}_n\approx d{\bf x}_n$,
the transformation would simply yield
$(d{\bf x})^2 = g_{\mu\nu} dq^\mu dq^\nu$. For
finite $\Delta {\bf x}_n$, however, we must expand $(\Delta {\bf x}_n)^2$ up to
forth order
in $\Delta {q_n}^\mu =  {q_n}^\mu - {q_{n-1}}^\mu$ since only
this yields all terms that will eventually contribute to
order $\epsilon$ \cite{eg,schmcl}. We expand around the final point $q_n^\mu $
(omitting for brevity the argument $q_n$ in
the ${e^i}_\mu$'s as well a the subscripts $n$ of $\Delta q^\mu$):
\begin{eqnarray} \label{23}
\lefteqn{{x^i (q_{n-1}) \equiv x^i (q_n - \Delta q_n)}=} \\
&&x^i(q_n)-{e^i}_\mu\Delta q^\mu +\frac{1}{2}
         {e^i}_{\mu ,\nu} \Delta q^\mu \Delta {q}^\nu
     -\frac{1}{3!} {e^i}_{\mu ,\nu\lambda} \Delta q^\mu
         \Delta q^\nu \Delta q^\lambda + \dots~. \nonumber
\end{eqnarray}

Squaring $\Delta {\bf x}_n$ and expressing
everything in terms of the
affine connection leads to the
short-time sliced action expressed entirely in terms of
intrinsic quantities (omitting again all sub $n$'s),
\begin{eqnarray} \label{24}
\lefteqn{
{\cal A}_>^\epsilon(q, q-\Delta q)  =  \frac{M}{2\epsilon}
\{ g_{\mu\nu} \Delta q^\mu \Delta q^\nu - \Gamma_{\mu\nu\lambda}
\Delta q^\mu \Delta q^\nu \Delta q^\lambda  }  \\
& & + \left [ \frac{1}{3} g_{\mu\tau} (\partial_\kappa
    {\Gamma_{\lambda\nu}}^\tau+
    {\Gamma_{\lambda\nu}}^\delta  {\Gamma_{\kappa\delta}}^\tau)
+ \frac{1}{4} {\Gamma_{\kappa\lambda}}^\sigma
\Gamma_{\nu\kappa\sigma}\right ]
\Delta q^\mu \Delta q^\nu \Delta q^\lambda \Delta q^\kappa+\dots \}
\nonumber
\end{eqnarray}
with $g_{\mu\nu}\equiv e^i{}_\mu e^i{}_\nu$ and $\Gamma_{\mu\nu\lambda}$
evaluated at the final
point $q$.
The measure of path integration in (\ref{21}) is
transformed to $ q^\mu $-space with a Jacobian
following from (\ref{23}),
\begin{equation} \label{25}
J = \frac{\partial(\Delta x)}{\partial(\Delta q)} =
\det(e^i{}_\kappa)\det(\delta^\kappa_{~~\mu} - {e_i}^\kappa e^i_{\{ \mu,\nu\}}
\Delta q^\nu + \frac{1}{2} e_i^\kappa e^i_{\{ \mu,\nu\lambda\} }
\Delta q^\nu \Delta q^\lambda + \dots),
\end{equation}
where the curly brackets around the indices denote their symmetrization.
Expanding the second
determinant in powers of $ \Delta q^\mu $, writing
$ \det({e^i}_\mu) \equiv e(q)
= \sqrt{\det g_{\mu\nu}(q)} \equiv \sqrt{g(q)}$, and
expressing the series in terms of a ``Jacobian effective action" $ {\cal A}_J$
with the definition
$ J=\sqrt{ g(q)}\exp\{i{\cal A}_J/\hbar \}$, we find
\begin{eqnarray} \label{26}
\frac{i}{\hbar}{\cal A}_J
&=&-\Gamma_{\{\nu\mu\}}{}^\mu \Delta q^\nu \\
& + & \frac{1}{2} [\partial_{\{\mu}
\Gamma_{\nu,\kappa\} }{}^\kappa + {\Gamma_{\{ \nu,\kappa}}^\sigma
{\Gamma_{\mu\},\sigma}}^\kappa - {\Gamma_{\{ \nu\kappa\} }}^\sigma
{\Gamma_{\{ \mu,\sigma\} }}^\kappa ] \Delta q^\nu \Delta q^\mu
+ \dots\nonumber
\end{eqnarray}
and arrive at the time-sliced path integral
in $q$-space
\begin{eqnarray} \label{27}
\lefteqn{\!\!\!\!\!\!\!\!\!\!\!\!\!\!\!\!\!\!\!\!\!\!
\langle q\vert \exp \{ - \frac{1}{\hbar}(t-t')\hat{H}\}\vert q'
\rangle \approx
\frac{1}{\sqrt{2\pi \hbar\epsilon/M}^D}\prod_{n=2}^{N+1}\left[
\int{d^D \Delta q_n}
 \frac{\sqrt{g(q_n)}}{\sqrt{2\pi \epsilon\hbar/M}^D}\right]  }\nonumber \\
&&\times \exp \left\{- \frac{1}{\hbar} \sum_{n=1}^{N+1}[{\cal A}_>^\epsilon
(q_n, q_n-\Delta q_n) +{\cal A}_J]\right\}
\end{eqnarray}
The integrals over $\Delta q_n$ are to be performed successively from $n=N$
down to $n=1$.

Our path integral is to be contrasted
with that of earlier works.
Expressed in our language, they start out from the
time-sliced flat-space path integral
\begin{equation} \label{21b}
({\bf x},t \vert {\bf x}',t') =
\frac{1}{\sqrt{2\pi \epsilon\hbar/M}^D}\prod_{n=2}^{N+1}
\left[
\int_{-\infty}^{\infty} dx_n \right] \prod_{n=1}^{N+1} K_0^\epsilon (\Delta
{\bf x}_n).
\end{equation}
In flat space, this is the same thing as (\ref{21}). Under
a non-holonomic transformation, however, it is mapped
into a {\em different\/} final expression. Here the measure
would go over into the naive group invariant measure and the amplitude
would read
\begin{eqnarray} \label{28}
\lefteqn{\!\!\!\!\!\!\!\!\!\!\!\!\!\!\!\!\!\!\!\!
\!\!\!\!\!\!\!\!\!\!\!\!\!\!
\langle q\vert \exp \{ - \frac{1}{\hbar}(t-t')\hat{H}\}\vert q'
\rangle \approx
\frac{1}{\sqrt{2\pi \hbar\epsilon/M}^D}\prod_{n=2}^{N+1}\left[
\int{d^D q_{n-1}}
 \frac{\sqrt{g(q_{n-1})}}{\sqrt{2\pi \epsilon\hbar/M}^D}\right]  }\nonumber \\
&&~~~~~~~\times \exp \left\{- \frac{1}{\hbar} \sum_{n=1}^{N+1}[{\cal
A}_>^\epsilon
(q_n, q_n-\Delta q_n)\right\}.
\end{eqnarray}
Expressing the correct amplitude (\ref{27}) in terms of
this naively expected measure we see that it reads
\begin{eqnarray} \label{28}
\lefteqn{\!\!\!\!\!\!\!\!\!\!\!\!\!\!\!\!\!\!\!\!
\langle q\vert \exp \{ - \frac{1}{\hbar}(t-t')\hat{H}\}\vert q'
\rangle \approx
\frac{1}{\sqrt{2\pi \hbar\epsilon/M}^D}\prod_{n=2}^{N+1}\left[
\int{d^D q_{n-1}}
 \frac{\sqrt{g(q_{n-1})}}{\sqrt{2\pi \epsilon\hbar/M}^D}\right]  }\nonumber \\
&&\times \exp \left\{- \frac{1}{\hbar} \sum_{n=1}^{N+1}[{\cal A}_>^\epsilon
(q_n, q_n-\Delta q_n) +\Delta_{meas} {\cal A}_J]\right\}
\end{eqnarray}
with a correction term $\Delta_{meas} {\cal A}_J$
which is the
difference
\begin{equation}
\Delta_{meas} {\cal A}_J= {\cal A}_J- {\cal A}_{J_0}
\label{jj}\end{equation}
between $ {\cal A}_J$ of (\ref{26}) and
$ {\cal A}_{J_0}$ that arises when bringing the measure in (\ref{27})
$$\prod_{n=2}^{N+1}\left[
\int{d^D \Delta q_n}
 \frac{\sqrt{g(q_n)}}{\sqrt{2\pi \epsilon\hbar/M}^D}\right]$$
 to the naively expected form in (\ref{28}),
i.e.
\begin{equation}
\exp\{i{\cal A}_{J_0}/\hbar \}\equiv    \sqrt{ g(q_n)}/\sqrt{
g(q_{n-1})}=e(q_n)/e(q_{n-1}).
\label{j0}\end{equation}
Expanding the determinant $ e(q_n)=e(q_{n-1}-\Delta q_n)$ in powers of
$ \Delta q_n$ we see that $ {\cal A}_{J_0}$ is the same as
$ {\cal A}_J$ in (\ref{26}) except that symmetrization symbols are absent:
\begin{eqnarray} \label{26b}
\frac{i}{\hbar}{\cal A}_{J_0}
&=&-\Gamma_{\{\nu\mu\}}{}^\mu \Delta q^\nu \\
& + & \frac{1}{2} [\partial_{\mu}
\Gamma_{\nu,\kappa }{}^\kappa + {\Gamma_{ \nu,\kappa}}^\sigma
{\Gamma_{\mu,\sigma}}^\kappa
- {\Gamma_{ \nu\kappa }}^\sigma
{\Gamma_{ \mu,\sigma }}^\kappa ] \Delta q^\nu \Delta q^\mu + \dots,\nonumber
\end{eqnarray}
Either (\ref{27}) or (\ref{28})
may be used as the correct path integral formulas
in spaces with curvature and torsion. \\~ \\
%
3)~As an application consider now the path integral
for the above discussed point particle
on the surface of a sphere in $D$ dimensions.
First we solve an auxiliary problem
{\em near\/} the surface of the sphere.
Its imaginary-time-sliced form reads
\begin{equation}\label{1}
({\bf  u}_b \tau _b \vert {\bf  u}_a \tau _a) \approx
\frac{1}{\sqrt{2\pi  \hbar \epsilon /Mr^2}^{D-1}}
\prod_{n=1}^N
\left[ \int \frac{d^{D-1}{\bf u}_n} {\sqrt{ 2\pi \hbar \epsilon /Mr^2}^{D-1}}
\right]
\exp\left\{ -\frac{1}{\hbar} {\cal A}^N\right\} ,\nonumber \\
\end{equation}
with the sliced action
\begin{equation} \label{2}
{\cal A}^N_0 = \frac{M}{2\epsilon }r^2 \sum_{n=1}^{N+1}({\bf u}_n -{\bf
u}_{n-1})^2
= \frac{M}{\epsilon }r^2 \sum_{n=1}^{N+1}(1-\cos{\Delta \vartheta} _n)
,
\end{equation}
where ${\Delta \vartheta} $ is the small angle between
${\bf u}_n$ and ${\bf u}_{n-1}$ (the $\approx $ sign becomes an equality for
$N\rightarrow {\infty} $).
There are two reasons for using the term {\em near\/} rather than {\em on\/}
the sphere.\\
i)~The
sliced action of the solvable path integral
(17) involves the shortest distances between the points in the {\em
embedding\/}
euclidean space rather than the intrinsic geodesic distances on
the sphere. This will be easy to correct
\cite{ji1,sch0}. \\
ii)~There is an additional action associated with the measure of
path integration given by (14) \cite{curtor}.

The  exact
solution of the auxiliary path integral (\ref{1}) goes as follows:
For each time interval $\epsilon $,
the exponential
$
 \exp \left\{ -{Mr^2}
(1- \cos {\Delta \vartheta}  _n)/\hbar \epsilon \right\}
$
is expanded into spherical harmonics according to formula
\begin{eqnarray} \label{3}
\lefteqn{
\exp \left\{  -\frac{Mr^2}{2\hbar \epsilon }({\bf u}_n
     -{\bf u}_{n-1})^2\right\}   ~~~~~~~~~~~~}\nonumber \\
& & =
     \sum_{l=0}^{\infty }a_l (h)
     \frac{l+D/2-1}{D/2-1} \frac{1}{S_D} C_l^{(D/2-1)}
         (\cos {\Delta \vartheta}  _n)\nonumber \\
&&~~~~~~~~~~~~~~~~~~~=   \sum_{l=0}^{\infty } a_l (h)
\sum _{\sbf m}Y_{l{\sbf m}}({\bf u}_n)
        Y^*_{l{\sbf m}} ({\bf u}_{n-1})
\end{eqnarray}
where
$
a_l(h) \equiv  \left( {2\pi }/{h}\right)^{(D-1)/2}
\tilde{I}_{l+D/2-1} (h)
$
,
$
\tilde I_\nu (z)\equiv \sqrt{ 2\pi z}e^{-z}I_\nu (z)
$ with $ I_\nu (z)=$ Bessel functions,
and
$
h\equiv  {Mr^2}/{\hbar \epsilon }
$.
The functions $ C^{(\nu )}_l(z )$ are the Gegenbauer polynomials and
$ Y_{l\sbf{m}}({\bf u})$ the hyperspherical harmonics in $ D$ dimensions
\cite{vi}.
For each adjacent pair of such factors
$(n+1,n), (n,n-1)$,
the integration over the intermediate ${\bf u}_n$
variable can be done using the well-known orthonormality relation
for the hyperspherical harmonics.
The combined two-step amplitude has the same expansion
as (\ref{3})
with $ a_l(h)$  replaced by $ (h/2\pi )a_l(h)^2$.
By successive integration in (\ref{1}) we obtain the
total time sliced amplitude
\begin{equation} \label{4}
({\bf u}_b \tau _b \vert {\bf u}_a \tau _a) \approx
      \left( \frac{h}{2\pi }\right)^{(N+1)(D-1)/2}
  \sum_{l=0}^{\infty } a_l(h)^{N+1}
\sum _{{\sbf m}}      Y_{l{\sbf m}} ({\bf u}_b) Y_{l{\sbf m}}^*({\bf u}_a) .
\end{equation}
We now go to the limit $N \rightarrow \infty ,
\epsilon  = (\tau _b-\tau _a)/(N+1) \rightarrow 0,$ where
\begin{eqnarray} \label{5}
&&\left( \frac{h}{2\pi }\right) ^{(N+1)(D-1)/2}
 a_l (h)^{N+1} =
    \left[  \tilde{I}_{l+D/2-1} \left( \frac{Mr^2}{\hbar \epsilon }
    \right)
    \right] ^{N+1}\nonumber  \\
& &~~~~~~~~ \longrightarrow   \exp \left\{ -
                   (\tau _b - \tau _a)\hbar  \frac{(l+D/2-1)^2-1/4}
                   {2Mr^2} \right\},
\end{eqnarray}
and obtain the time displacement amplitude for
the motion {\em near\/} the sphere as the spectral expansion
\begin{equation} \label{6}
({\bf u}_b \tau _b \vert {\bf u}_a \tau _a) =
 \sum_{l=0}^{\infty } \exp \left\{ -
         \frac{\hbar {L}_2}{2Mr^2}(\tau _b-\tau _a) \right\}
\sum _{{\sbf m}} Y_{l{\sbf m}} ({\bf u}_b) Y_{l{\sbf m}}^{*}({\bf u}_a),
\end{equation}
with
\begin{equation} \label{7}
{L}_2 \equiv  (l+D/2-1)^2 -{1}/{4},
\end{equation}
and the energy eigenvalues $ E_l^{near}=\hbar ^2L_2/2Mr^2$.
For $ D=4$, the most convenient expansion is in terms of the representation
functions $ {\cal D}_{mm'}^l(\varphi ,\theta ,\gamma )$ of the
rotation group, involving the Euler angle parametrization of
the
vectors on the unit sphere
\begin{eqnarray} \label{8}
\hat x^1 & = &\cos(\theta /2)\cos[(\varphi +\gamma )/2] \nonumber \\{}
\hat x^2 & = &-\sin(\theta /2)\sin[(\varphi +\gamma )/2] \nonumber \\{}
\hat x^3 & = &\sin(\theta /2)\cos[(\varphi -\gamma )/2] \nonumber \\{}
\hat x^4 & = &\sin(\theta /2)\sin[(\varphi -\gamma )/2].
\end{eqnarray}
In terms of these,
(\ref{6}) reads
\begin{eqnarray} \label{9}
\lefteqn{\!\!\!\!\!\!\!\!\!\!\!\!\!\!\!\!\!
({\bf u}_b \tau _b \vert {\bf u}_a \tau _a) =
 \sum_{l=0}^{\infty } \exp \left\{ -
         \frac{\hbar {L}_2}{2Mr^2}(\tau _b-\tau _a) \right\}}\nonumber \\
&&\times\sum _{m_1,m_2=-l}^l \frac{l+1}{2\pi ^2}{\cal D}^{l/2}_{m_1m_2}
(\varphi _b,\theta _b,\gamma _b)
{\cal D}^{l/2~~*}_{m_1m_2}(\varphi _{a},\theta _{a},\gamma _{a}).
\end{eqnarray}
These amplitudes display the correct
wave functions for the movement {\em on\/} the surface of the sphere,
as we know from Schr\"odinger theory.
They do not, however, carry the correct energy eigenvalues
which should be
$ E_l=\hbar ^2\hat L^2/2Mr^2$ with the eigenvalue of the squared angular
momentum operator
$ \hat L^2=l(l+D-2)$ rather
than $ E_l^{near}$ with ${L}_2 = (l+D/2-1)^2 -{1}/{4}$.

To have the correct energies, the
path integral needs the two changes announced above.
First, the time-sliced action must
measure the proper geodesic distance
rather than the
euclidean distance in the embedding space and should thus read
\begin{equation} \label{10}
{\cal A}^N = \frac{M}{\epsilon }r^2 \sum_{n=1}^{N+1}
\frac{( {\Delta \vartheta}  _n)^2}{2},
\end{equation}
rather than (\ref{2}).
Since the time-sliced path integral is solved exactly
with the action (\ref{2}) it is convenient to
expand the true action around the soluble one as \cite{r2}
\begin{equation} \label{11}
{\cal A}^N ={\cal A}^N_0+\Delta_4 {\cal A}^N= \frac{M}{\epsilon }r^2
\sum_{n=1}^{N+1}
\left[ {(1-\cos {\Delta \vartheta}  _n)}+\frac{1}{24}{\Delta \vartheta}
_n^4+\dots\right]  ,
\end{equation}
and treat the correction perturbatively to lowest order.
There is no need to go higher than quartic order since
only the quartic term contributes to the relevant order $ \epsilon $
in the limit $N\rightarrow {\infty} $.
In $ D=2$  dimensions, the quartic correction is sufficient to bring the path
integral
from {\em near\/} to {\em on\/} the sphere (here a circle).
Indeed, with the measure of the path integration being
\begin{equation} \label{12}
\frac{1}{\sqrt{ 2\pi \hbar \epsilon /Mr^2}}\prod _{n=1}^{N+1}\int_{-\pi
/2}^{\pi /2}\frac{d\varphi _n}{\sqrt{ 2\pi \hbar \epsilon /M
r^2}}
\end{equation}
and the leading action (\ref{10}),
the quartic term
$ {\Delta \vartheta} _n^4=(\varphi _n-\varphi _{n-1})^4$ can be replaced
by its expectation
\begin{equation} \label{13}
\langle {\Delta \vartheta} _n^4\rangle _0=3\frac{\epsilon \hbar }{Mr^2},
\end{equation}
so that the correction term of the action
is given by
\begin{equation} \label{14}
\langle \Delta_4 {\cal A}^N\rangle _0=(\tau _b-\tau _a) \frac{\hbar
^2/4}{2Mr^2}.
\end{equation}
where we have replaced $(N+1)\epsilon $ by $\tau _b-\tau _a$.
For $ D=2$, this supplies precisely the missing energy to raise $ E_l^{near}$
up to
$ E_l$.

In higher dimensions, we must change also the measure of path integration
is necessary according to \cite{curtor}.
What we have to explain in any $D$ is the difference
\begin{equation} \label{15}
\Delta L_2=\hat L^2-L_2 =1/4-( D/2-1)^2.
\end{equation}
This vanishes at $ D=3$ where it changes sign. Note that
the expectation of the
quartic correction term $\Delta_4 {\cal A}^N$
in (\ref{11}) being always positive
cannot account for the discrepancy by itself.
Let us calculate its contribution in $ D$  dimensions. For very small $
\epsilon $, the fluctuations
near the sphere will lie close to the $ D-1$ dimensional tangent
space. Let $ \Delta {\bf x}_n$ be the coordinates in this space.
Then we can write
\begin{equation} \label{16}
\Delta_4 {\cal A}^N \approx  \frac{M}{\epsilon }r^2 \sum_{n=1}^{N+1}
\frac{1}{24}( \frac{\Delta {\bf x}_n}{r})^4.
\end{equation}
The $ \Delta {\bf x}_n$'s have the lowest order correlation
$
\langle \Delta { x_i}\Delta x_j\rangle _0=({\hbar \epsilon }/{M})\delta _{ij}
$.
This shows that $ \Delta_4 {\cal A}^N$
has the expectation
\begin{equation} \label{17}
\langle \Delta {\cal A}^N\rangle _0=(\tau _b-\tau _a)\frac{\hbar ^2}{2Mr^2}
 \Delta_4 L_2
{}.
\end{equation}
where $ \Delta_{4} L_2$ is the contribution of the quartic term to the
value $ L_2$,
\begin{equation} \label{18}
\Delta_{4} L_2=\frac{D^2-1}{12}.
\end{equation}
This result is obtained using the Wick contraction rules for the tensor
$$ %
 \langle \Delta x_i\Delta x_j\Delta x_k\Delta x_l\rangle _0
=({\epsilon \hbar }/{M})(\delta _{ij}\delta _{kl}+\delta _{ik}\delta _{jl}+
\delta _{il}\delta _{jk}).
$$
Thus we remain with a final discrepancy in $ D$ dimensions,
\begin{equation} \label{19}
\Delta_{meas} L_2=\Delta L_2-\Delta_{4} L_2 =-\frac{1}{3}(D-1)(D-2),
\end{equation}
to be explained now.

The present problem involves no torsion. Then a simple algebra
shows that $ \Delta_{meas} {\cal A}_J$ defined by (\ref{jj})
with (\ref{26}), (\ref{26b}) reduces to
\begin{equation} \label{29}
 \Delta_{meas} {\cal A}_J=-\frac{\hbar }{6}\bar R_{\mu \nu }\Delta q^\mu \Delta
q^\nu ,
\end{equation}
where $ \bar R_{\mu \nu }$
is the Ricci tensor, which for a sphere of radius $r$ is \linebreak[4]
$ (D-2)g_{\mu \nu }/r^2$. The perturbative treatment of (\ref{29})
gives the only relevant contribution to the energy,
\begin{equation} \label{30}
\langle  \Delta_{meas} {\cal A}_J\rangle _0 =-\epsilon \frac{\hbar
^2}{6M}\frac{(D-1)(D-2)}{r^2},
\end{equation}
thus producing precisely the missing energy required by (\ref{19}).\\ ~ \\
4)~The
sphere in four dimensions is equivalent to the covering
group of rotations in three dimensions, the group $ SU(2)$.
Knowing now how to solve the time-sliced
path integral
near and on the surface of the sphere, we can obtain the same quantities
{\em near\/}  and {\em on\/} the group space of $ SU(2)$ \cite{bj}.
This puts us in a position to solve the time sliced path integral of
a spinning spherical top by reduction to the
$ SU(2)$ problem. We only have to go from $ SU(2)$,
which is the covering group
of the rotation
group, down to the rotation group itself \cite{r4}.
The angular positions with Euler angles $ \gamma $ and $ \gamma +2\pi $
are physically indistinguishable.
The physical states must be a representation of this operation and
the time-displacement amplitude must reflect this.
The simplest possibility is the trivial even representation
where one adds the amplitudes to go from the
initial configuration $ \varphi _a,\theta _a,\gamma _a$ to
the identical final ones $ \varphi _b,\theta _b,\gamma _b$
and $ \varphi _b,\theta _b,\gamma _b+2\pi $ and forms the
amplitude
\begin{eqnarray} \label{31}
\lefteqn{
( \varphi _b,\theta _b,\gamma _b~\tau _b| \varphi _b,\theta _b,\gamma _b~\tau
_a)_{top} =}\nonumber \\
&&( \varphi _b,\theta _b,\gamma _b~\tau _b| \varphi _b,\theta _b,\gamma _b~\tau
_a)
+( \varphi _b,\theta _b,\gamma _b+2\pi ~\tau _b| \varphi _b,\theta _b,\gamma
_b~\tau _a).
\end{eqnarray}
The sum eliminates all half-integer
representation functions ${ d}^{l/2}_{mm'}(\theta )$
in the expansion (\ref{1}) of the amplitude.

Instead of the sum we could also have formed another
representation of the operation $ \gamma \rightarrow \gamma +2\pi $, the
antisymmetric combination
\begin{eqnarray} \label{32}
\lefteqn{
( \varphi _b,\theta _b,\gamma _b~\tau _b| \varphi _b,\theta _b,\gamma _b~\tau
_a)_{fermions} =}\nonumber \\
&&( \varphi _b,\theta _b,\gamma _b~\tau _b| \varphi _b,\theta _b,\gamma _b~\tau
_a)
-( \varphi _b,\theta _b,\gamma _b+2\pi ~\tau _b| \varphi _b,\theta _b,\gamma
_b~\tau _a).
\end{eqnarray}
Here the expansion (\ref{9}) retains
only the half-integer angular momenta $ l/2$.
In nature such spins are associated with fermions such as electrons,
protons, muons, or neutrinos, which carry only one specific value of $l/2$.

In principle, there is no problem in treating also a non-spherical top.
While the spherical top has ``near the group space" a
time-sliced
action
\begin{equation} \label{33}
\frac{1}{\epsilon ^2}I\left[ 1-\frac{1}{2}\mbox{tr}(g_ng_{n-1}^{~-1})\right] ,
\end{equation}
the asymmetric top with three moments
of inertia $I_{123}$ requires separating the three components of the
angular velocities
\begin{equation} \label{34}
\omega _a={i}\mbox{tr}(g\sigma_a\dot {g^{-1}}),~~~a=1,2,3 .
\end{equation}
($\sigma _a$= Pauli matrices)
on the time lattice
so that the action reads
\begin{eqnarray} \label{35}
&&\frac{1}{\epsilon ^2}\left\{ I_1   [1-\frac{1}{2}\mbox{tr}(g_n\sigma _1
g_{n-1}^{~-1})]
+I_2  [1-\frac{1}{2}\mbox{tr}(g_n\sigma _2  g_{n-1}^{~-1})]
\right .\nonumber \\
&&\left .~~~~~~~~~~~~~~+I_3   [1-\frac{1}{2}\mbox{tr}(g_n\sigma _3
g_{n-1}^{~-1})]\right \}.
\end{eqnarray}
rather than (\ref{33}).
The amplitude ``near the top" is then an
appropriate generalization of (\ref{9}).
The calculation of the correction term
$ \Delta E$, however, is more complicated
than before and is left to the reader, following the rules explained above.\\
{}~\\
5)~As mentioned above, the correctness of
the torsional aspects of the
proposed path integral (\ref{27}) or (\ref{28})
can be tested by
applying it to the path integral of the Coulomb system.
The procedure is too lengthy to be presented here and we refer the
reader to the textbook \cite{pi} for a detailed discussion.

%


\newpage
\begin{thebibliography}{11}
\bibitem{hol}
We call a coordinate transformation  $dx^i\rightarrow e^i_\mu (q)dq^\mu $
{\em non-holonomic\/} if either $\partial _\mu e^i_\nu -\partial _\mu e^i_\nu
\neq0$
or $(\partial _\mu \partial _\nu -\partial _\nu \partial _\mu )e^i_\lambda
\neq0$.
In the first case, the mapping carries a flat space into one
with torsion, in the second case to one with curvature.

\bibitem{curtor}
H. Kleinert,
Mod. Phys. Lett. A {\em 4\/}, 2329 (1989);
Phys. Lett B {\em 236\/}, 315 (1990);

\bibitem{pi}
H. Kleinert, {\em Path Integrals in Quantum Mechanics, Statistics,
and Polymer Physics\/}, World Scientific, Singapore, 1995.

\bibitem{dk}
I.H. Duru and H. Kleinert,
Phys. Letters {\em B 84\/}, 30 (1979);
Fortschr. Physik {\em 30\/}, 401(1982).


\bibitem{dw}
B.S. DeWitt, Rev. Mod. Phys. {\em 29\/}, 377 (1967)

\bibitem{ch}
K.S.~Cheng, J. Math. Phys. {\em 13\/}, 1723 (1972).

\bibitem{gav}
For other discussions with results different from
the above and ours see\newline
H. Kamo and T. Kawai, Prog. Theor. Phys. {\em 50\/}, 680, (1973);\\
T. Kawai, Found. Phys. {\em 5\/}, 143 (1975);\\
H. Dekker, Physica {\em 103A\/}, 586 (198);\\
G.M. Gavazzi, Nuovo Cimento {\em A~101\/}, 241 (1981).

\bibitem{howtodo}
H. Kleinert,
Phys. Lett. {\em A120\/}, 361 (1987)




\bibitem{hi}
There was an earlier claim in the literature
to have solved
the $D=3$ time slicing problem in the path integral of the Coulomb system
by R. Ho and A. Inomata, Phys. Rev. Lett.
{\em 48\/}, 231 (1982). They, however, start from
Feynman's path integral formula and
do not proceed consistently, arriving
at the correct (known) final result
only
thanks to an inconsistent treatment of the measure. See \cite{howtodo}. We know
now
that it is impossible to use the Feynman formula as a starting place, due
to path collapse. See
H. Kleinert, Phys. Lett. {\em B224\/}, 313 (1989). This is why also
the approach
taken by
F. Steiner, Phys. Lett. {\em 106A\/}, 356, 363 (1984)
is incorrect.


\bibitem{gfcm}
The construction of such mappings is standard
in the theory of defects in crystals, see\\
H. Kleinert, {\em Gauge Fields in Condensed Matter\/}, Vol. II, World
Scientific,
Singapore 1989.




\bibitem{eg}
S. Edwards, Y. Gulyaev, Proc. Roy. Soc. London, {\em A 279\/}, 229 (1964).


\bibitem{schmcl}
D.W. Mc. Laughlin, L.S. Schulman, J. Math. Phys. {\em 12\/}. 2520 (1971)


\bibitem{ji1}
G. Junker and A. Inomata,
in {\em Path Integrals from meV to MeV\/}, ed. by M. C. Gutzwiller et al.,
World Scientific 1986

\bibitem{sch0}
L. Schulman, Phys. Rev. {\em 174\/}, 1558 (1968).




\bibitem{vi}
H. Bateman, {\em Higher Transcendental Functions\/}, McGraw-Hill, New York,
1953, Vol II, Ch. XI and \newline N.H. Vilenkin,
{\em Special Functions and the Theory of Group Representations\/}, Am. Math.
Soc.,
Providence, R I, 1968.

\bibitem{r2}
This step was still done in \cite{ji1}. Note,
however,  that the (known) correct
final result stated
in that paper is impossible to obtain from their calculation since
the measure problem, which
is the main issue of the present paper,
was not solved at that time.

\bibitem{bj}
For path integrals near group spaces (although claimed to
work on group spaces) see
M. B\"ohm, A. Junker, J. Math. Phys. {\em 30\/}, 1195 (1989).


\bibitem{r4}
This point was discussed by
L. Schulman, Phys. Rev. {\em 174\/}, 1558 (1968).
The author also gives a correct path integral but he does so a posteriori, by
reconstructing it from the known spectral representation of the
Schr\"odinger result.


\end{thebibliography}
\end{document}